\documentclass[12pt]{article}
\usepackage{latexsym}
\usepackage{epsfig}
\begin{document}

 \rightline{EFI 97-49}

\begin{center} \large {\bf Discrete Scale Invariance and the ``Second Black Monday"}

 \medskip\normalsize 
James A. Feigenbaum and Peter G.O. Freund \\

             {\em Enrico Fermi Institute and Department of Physics\\
             The University of Chicago, Chicago, IL 60637, USA}

 \end{center}
 \bigskip

\noindent{\bf Abstract}: Evidence is offered for log-periodic (in time) 
fluctuations in the S\&P 500 stock index during the three years 
prior to the October 27, 1997 ``correction''.
These fluctuations were expected on the basis of a discretely scale invariant
rupture phenomenology of stock market crashes proposed earlier. 
 
\bigskip

We \cite{FF} and others \cite{S, SJ} have proposed a picture of 
stock market ``crashes" as rupture processes with an underlying discrete scale 
invariance. The crash is viewed as a critical point in a system with discrete
scale invariance. On account of the appearance of complex critical exponents, 
such a picture predicts 
log-periodic (in time) fluctuations in stock market
indices (such as the S\&P 500) prior to the crash. In \cite{FF, S, SJ}
it was shown that such
fluctuations have indeed preceded the 1987 and 1929 crashes. Typically these 
fluctuations are observed over a period of years. With this in mind, we want 
to analyze in this brief note 
the S\&P 500 index during the period 1994-1997 prior to the 
October 27, 1997 ``correction". 

\begin{figure}[h,t]
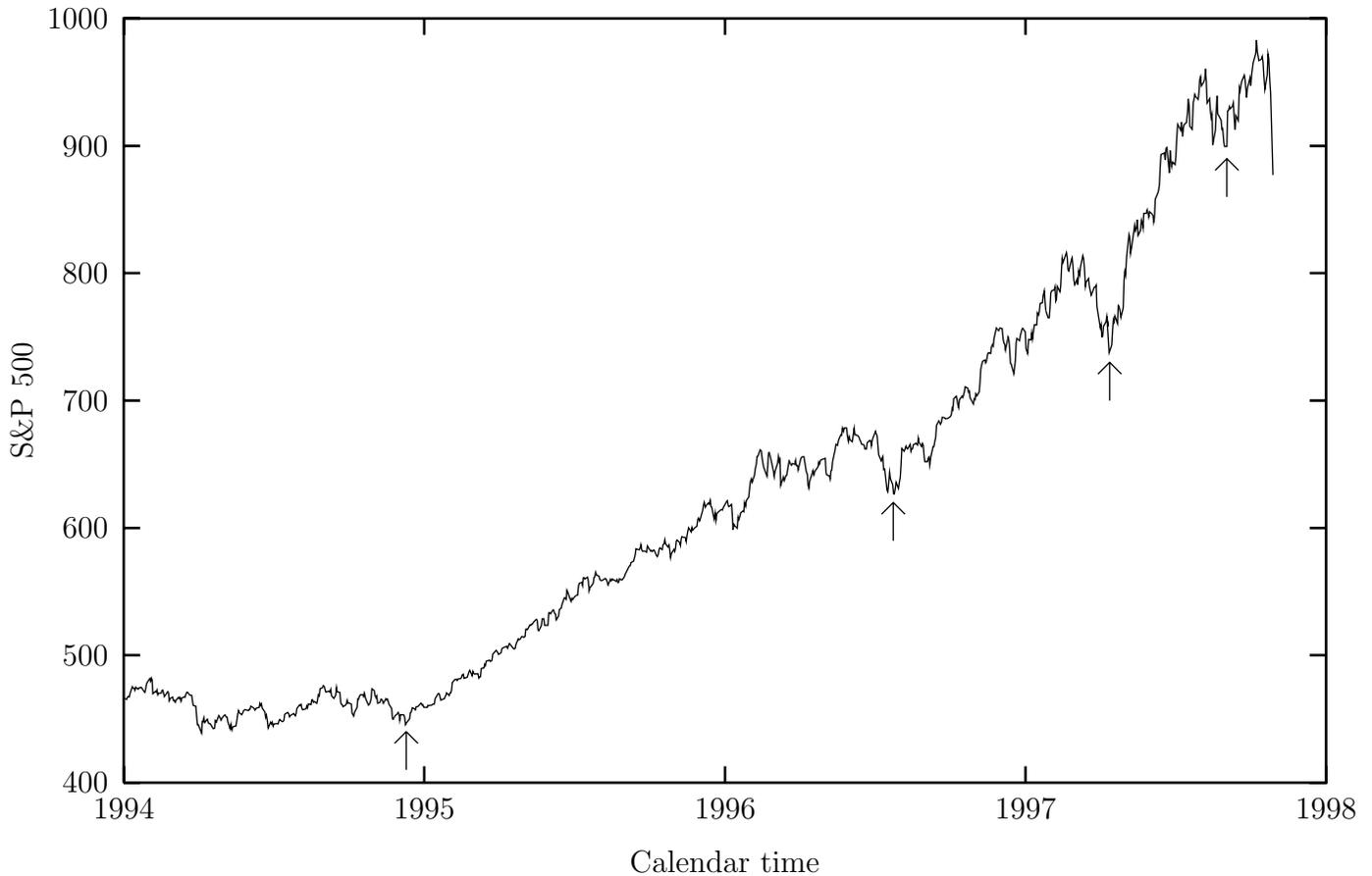

\include{sp}
\caption{Plot of the S\&P 500 index. The arrows indicate the troughs of the 
apparent log-periodic fluctuations.}
\end{figure}

In Fig. 1 we present the daily S\&P 500 data for this period 
and wish to draw attention to the typical log-periodic pattern 
pointed out with arrows at the troughs. Unlike the previous crashes 
analyzed in our earlier paper, the shape of these oscillations is not
harmonic, but rather of a ``sawtooth" type. Of course this is still
in agreement with discrete scale invariance, which requires log-periodicity 
but does not predict the shape of the periodic function. Correspondingly we 
make the Ansatz
\begin{equation}
c(t) = A + B (t_c - t)^{\alpha} [1 + (t_c - t)^{\beta} s(\ln (t_c - t))].
\end{equation}
Here $t$ is time, $t_c$ is the time of the ``correction", and
$c(t)$ is the S\&P 500
index. $A$, $B$, $\alpha$ and $\beta$ are constants and finally 
$s$ is a periodic
function of its argument. In our previous work $s$ was a harmonic function and 
$\beta$ was set equal to zero.

\begin{figure}[p]
\begin{picture}(450,500)(50,50)
\put(-50,0){\epsfig{file=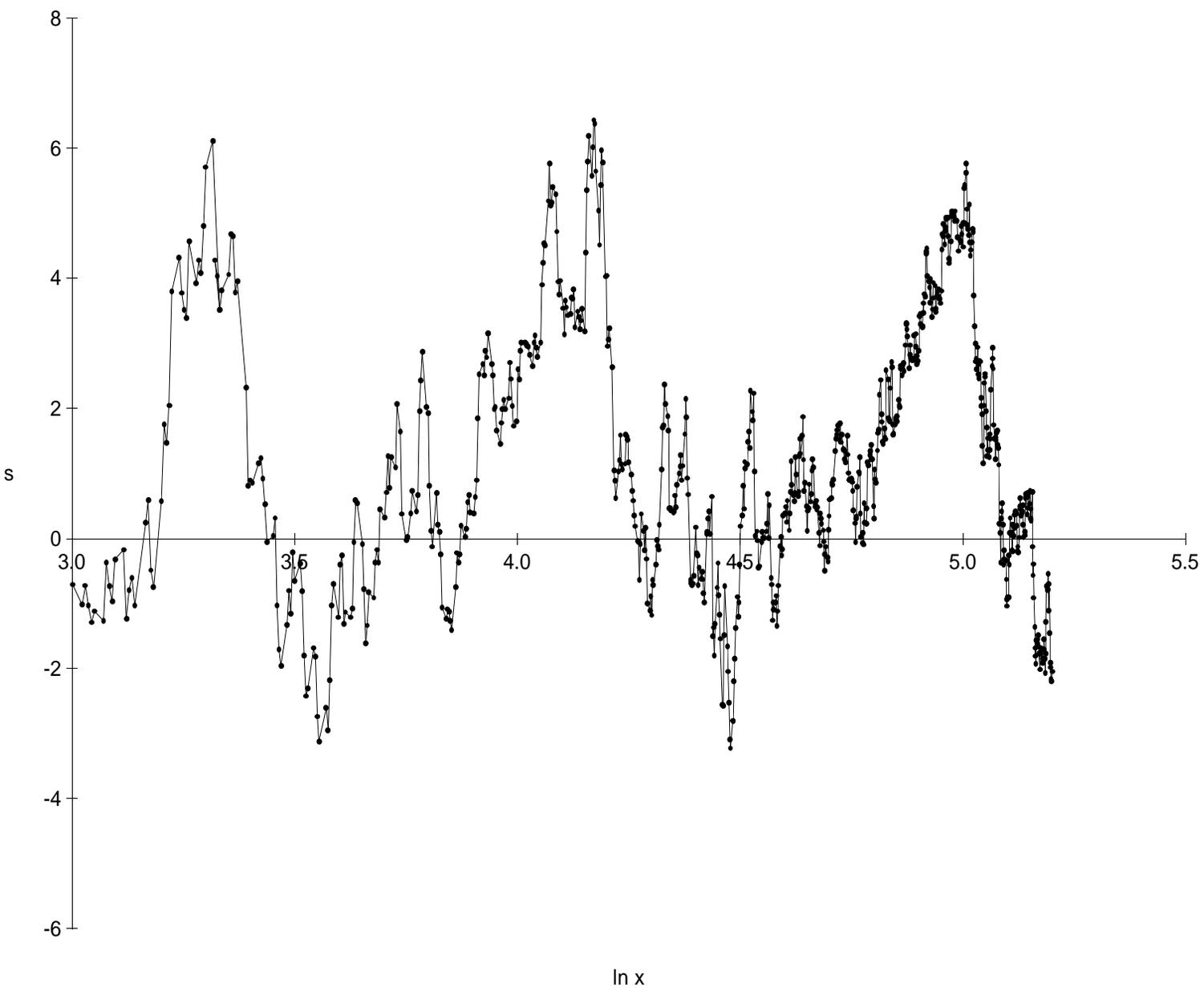, width = 8in, height=10in, angle=0}}
\end{picture}
\caption{The periodic function $s(\ln x)$, determined as explained in the 
text.}
\end{figure}

If the discrete-scaling hypothesis applies in this case, then
there should exist a 
choice of the parameters $A$, $B$, $\alpha$, and $\beta$ such that
the function $s$ calculated from the above equation should be periodic
in $\ln x$ where $x=t_c-t$. For the choice $A = 972.28$, $t_c = $10/21/97 
(the date of the last peak before the ``correction"), $B = 20.26$,
$\alpha = 0.63059$, and $\beta = -0.8$, we obtain the function $s(\ln x)$
shown in Fig. 2. The log-periodic structure noted heuristically in Fig. 1 
is reflected in the periodicity visible in Fig. 2. 
The period can be read off this figure as  $\Delta \approx 0.8$ and 
corresponds to a frequency $\omega =2\pi/\Delta \approx 7.8$. This 
is comparable to the values of $\omega$, $8.88$ and $8.73$, read off 
our earlier fits for the 1987 and 1929 crashes. The rightmost
period in Fig. 2 corresponds to the time interval between the arrows in the 
years 1994 and 1996 in Fig. 1. The middle period in Fig. 2 corresponds to the 
time interval 
between the 1996 arrow and the early 1997 arrow in Fig. 1. The leftmost 
portion of Fig. 2
corresponds to part of the much narrower interval between the two 1997
arrows in Fig. 1, This portion is close to the October event and, as in past 
fits \cite{FF, S, SJ}, the statistical fluctuations are larger in this region.

Monitoring log-periodic oscillations in 
the index over the three year period preceding the recent 1997
correction clearly indicated the impending rupture event. Whether and how
these oscillations are related to the time-evolution of events -- e.g. the
Asian currency devaluations -- currently believed to have caused this 
correction is an interesting open problem. Such events may have 
provided the small perturbation which unleashed the rupture signaled 
by the log-periodic oscillation pattern.

We wish to thank Mircea Pigli for his interest in this work. 
This work was supported in part by NSF Grant No. PHY-9123780-A3.

\end{document}